\documentclass[a4paper,fleqn,usenatbib,useAMS]{mnras}
\usepackage{graphicx}	
\usepackage{amsmath}	
\usepackage{amssymb}	
\usepackage{multicol}        
\usepackage{bm}		
\usepackage{pdflscape}	
\usepackage[T1]{fontenc}
\usepackage{ae,aecompl}

\title {{Effects of   Cosmological Constant  on  Clustering of Galaxies}}
\author[M.  Hameeda et al.]{Mir  Hameeda,$^{1}$\thanks{E-mail: hme123eda@gmail.com}
 Sudhaker Upadhyay,$^{2}$\thanks{E-mail: sudhakerupadhyay@gmail.com}  
 Mir Faizal$^{3}$\thanks{E-mail: mirfaizalmir@googlemail.com}  
and  Ahmed Farag Ali$^{4}$\thanks{E-mail: ahmed.ali@fsc.bu.edu.eg}  
\\
$^{1}$Department of Physics 
 S.P. Collage,  Srinagar, Kashmir, 190001, India \&\\
 Inter University Center for Astronomy and Astrophysics, IUCAA,  Pune-411007, India\\
$^{2}$Centre for Theoretical Studies, 
Indian Institute of Technology Kharagpur,  Kharagpur-721302,  India\\
$^{3}$Irving K. Barber School of Arts and Sciences, University of British Columbia - Okanagan,\\  3333 University Way, Kelowna,   British Columbia V1V 1V7, Canada \&\\ 
Department of Physics and Astronomy, University of Lethbridge,
Lethbridge, Alberta, T1K 3M4, Canada \\
$^{4}$Netherlands Institute for Advanced Study
Korte Spinhuissteeg 3, 1012 CG Amsterdam, Netherlands \&\\
Department  of Physics, Faculty of Science,  
 Benha University, Benha, 13518, Egypt} 
 
\begin{document}
 
\maketitle

\begin{abstract}
In this paper, we analyse the effect of the expansion of the universe on the clustering of galaxies. 
We evaluate the configurational integral for interacting system of galaxies in an expanding universe by   including effects 
produced by the cosmological 
constant. The  gravitational partition function  is obtained using this 
configuration integral.  
Thermodynamic quantities, specifically, Helmholtz free
energy, entropy, internal energy, pressure   and chemical potential are also  
 derived for this system.  It is observed that they   depend  on the modified clustering parameter for this system of galaxies. 
  It is also demonstrated that these thermodynamical quantities get corrected because of the cosmological constant. 
\end{abstract}

\begingroup
\let\clearpage\relax
\tableofcontents
\endgroup
\newpage
\section{Overview and Motivation} 
It has been establised that the Universe has an acceleration in its expansion, and this is based on the observations from the  
Type Ia Supernovae (SNeIa) \citet{[17]}. Thus, there seems to be some dark energy contribution to the 
total energy density of the Universe. 
Furthmore, the observations from 
the  anisotropies in the cosmic microwave background radiation (CMBR) \citet{[18]},
and  inferred  matter power spectrum from large galaxy surveys \citet{[19]} can also been used to make similar predications. 
According to the cosmological constant  $\Lambda$ cold dark matter ($\Lambda$CDM) model, the baryons contribute only for  
$\sim 4 \%$, while the exotic cold dark matter
(CDM) represents the bulk of the matter  contributes $\sim  25 \% $  and the cosmological constant  $\Lambda$ plays the role of the so
called ``dark energy”  contributes $ \sim 70 \%$  of the universe \citet{[20]}. 
It is expected that the general relativity might get corrected as 
 its validity on the larger  and  smaller scales has not been verified 
\cite{[24]}. Therefore, to explain the   cosmic speed up and dark matter, it is  possibile to generalize the Hilbert-Einstein Lagrangian, which is 
linear in the Ricci scalar $R$.  This is the basic idea behind the study of $f(R)$ theories of gravity 
\citep{[25],[26],[27],[28],[29],[30],[31]}.  It may be also noted that 
   various models of dark energy have been proposed to explain the late-time cosmic acceleration
without the cosmological constant. For instance, these models are
 a non-canonical scalar field such as phantom \citet{20}, tachyon scalar
field motivated by string theories \citet{21},   a fluid with a special equation of state  called
as Chaplygin gas \citet{22,23,24} and   a canonical scalar field,
so-called quintessence \citet{19}. In fact, some proposals of holographic dark energy have also been proposed \citet{25,26,27}.
The   $\Lambda$CDM  model, in which  dark energy is well-represented by  $\Lambda$ in Einstein's gravity, and it  is  also supported by various
cosmological observations.  So, despite of  several complications of baryonic astrophysics and lack of understanding behind the 
theoretical origin of the cosmological constant $\Lambda$  \cite{55},  the $\Lambda$CDM 
model is considered as the standard cosmological model describing  the Universe on large scales.

It is also possible to  analyse the effect of dark energy in the form of cosmological constant directly 
on the partition function of galaxies.  In this analysis, the galaxies can be approximated as point particles interacting 
through the Newton's law, and the effect of dark energy can be introduced as a additional correction term in this partition 
function.
In this approximation, we have a system of galaxies which can be approximated by a system of particles 
for our analysis, and thus formalism of statistical mechanics can be used for analysis this model \cite{ahm02}. 
Thus, it is possible to  study the  thermodynamics  of  interacting system of galaxies in the 
 expanding universe from the using the formalism of statistical mechanics. 
The formalism of statistical mechanical has been used to   analyse the observed peculiar velocity distribution function
for a   sample of galaxies within $50 Mpc (H=100)$ of the Local Group \cite{1a}. 
In this study, a   a wide range of
 clustering properties for this sample of galaxies have been analysed. 
In this study, the effects of uncertainties in sampling   on the estimated distribution function have also been studied. 
The peculiar velocity distribution function of galaxies has been used to analyse  a system of  galaxies with haloes
\cite{2a}. In this study, it  was  demonstrated that  individual massive galaxies are usually surrounded by their own haloes, 
 and they are not embedded in common haloes. This was done by comparing this study with the observed peculiar velocity 
 distributions. 
 
 The   spatial distribution function of galaxies at high redshift have also been analysed using the formalism of 
 statistical mechanics \cite{4a}. It has been demonstrated that 
 the redshifts of the galaxy spatial distribution function   has the same form as  predicted by gravitational quasi-equilibrium dynamics.
 This observation constrains the  processes such as merging of galaxy. 
 The    probability that a galaxy cluster of a given shape exists has also been analysed using the formalism of statistical mechanics 
 \cite{5a}. This has been done using the observation that 
 the distribution of galaxies is very close to quasi-equilibrium. This hold for  both its linear and nonlinear regimes.
 The   statistical mechanical formalism  of cosmological many-body problem  has also been used to analyse 
 a system   of two different  kind of  galaxies \cite{6a}. The general partition function for such a system has been obtained
 in the grand canonical ensemble.  This has been used to obtain various thermodynamical quantities for such a system of galaxies. 
 It has also been demonstrated that a softening parameter can be introduced in the partition function for galaxies, 
 if the finite size of galaxies is taken into account \cite{7a}. Thus, it is well establised that the formalism of statistical mechanics 
 can be used for analysing the clustering of galaxies.

 In this paper, we  analyse the effect of the cosmological constant on the clustering of galaxies.
This analysis is accomplished by deriving the gravitational partition function for galaxies 
in a universe with a    cosmological constant. We utilize here the configuration integrals over a spherical volume.  
The exact equations of state for galaxies  is also obtained, 
by computing Helmholtz free energy, entropy, internal energy, pressure and chemical potential, which depend on 
the corrected clustering parameter explicitly.  The corrections to the clustering parameter
are analysed for both point mass and non-point mass particles.
We investigate the effect of the 
 cosmological constant on the   distribution function for galaxies.
  We analyse the effect of approximating a galaxy as 
 a point mass and as an extended structure.  Thus, we are able to analyse  
  the effect of dark energy on the clustering of galaxies.

The organization of this paper is as follows. In section II, we elucidate the 
partition function for a system of galaxies with a  cosmological constant. The various thermodynamic quantities get corrected,  and such  corrections are derived in section III.
The details for general distribution functions are reported in section IV. 
In this  last section  we also summarize our results.   
 
\section{Gravitational Partition Function}
The a system of galaxies can be approximated as particle with  pairwise interaction. It is assumed that the distribution is
statistically homogeneous over large regions. 
The general partition function of a system of $N$ such galaxies  of mass $m$ interacting gravitationally with a potential energy
$\Phi$, having momenta $p_{i}$ and average temperature $T$ \citet{ahm02}:
\begin{eqnarray}
&&Z(T,V)=\frac{1}{\lambda^{3N}N!}\int d^{3N}pd^{3N}r \times\nonumber\\
&& \exp\biggl(-\biggl[\sum_{i=1}^{N}\frac{p_{i}^2}{2m}+\Phi(r_{1}, r_{2}, r_{3}, \dots, r_{N})\biggr]T^{-1}\biggr),
\end{eqnarray}
here $N!$ takes the distinguish-ability of classical particles into account, and $\lambda$ refers 
the normalization factor  resulting from integration over momentum space. 

Performing integration over momentum space yields,  
\begin{equation}
Z_N(T,V)=\frac{1}{N!}\left(\frac{2\pi mT}{\lambda^2}\right)^{3N/2}Q_N(T,V), \label{zn}
\end{equation}
where the configurational integral, $Q_{N}(T,V)$, is  given by 
\begin{equation}
Q_{N}(T,V)=\int....\int \prod_{1\le i<j\le N} exp[-\phi(r_{ij} )T^{-1}]d^{3N}r. \label{q1}
\end{equation}
In general, the gravitational potential energy, given as,
\begin{equation}
\sum_{1\le i<j\le N}\phi(r_{ij})= \Phi(r_{ij})= \Phi(r_{1}, r_{2}, \dots, r_{N}),
\end{equation} 
 is a function of the relative position vector $r_{ij}=|r_{i}-r_{j}|$ and is the sum of the potential energies of all pairs. In a gravitational system,  the potential energy $\Phi(r_{1}, r_{2}, \dots, r_{N})$ is due to all pairs of particles composing the system.

Here,  it is worth defining  two particle function by
\begin{equation}
f_{ij}=e^{-\Phi(r_{ij})/T}-1, \label{fun}
\end{equation}
which  appears only if the interactions are present in the system, however, it disappears in 
absence of interactions.  

It has been shown by   \citet{ahm02} that the configurational integral can be expressed as
\begin{eqnarray}
Q_{N}(T,V)&=&\int....\int \biggl[(1+f_{12})(1+f_{13})(1+f_{23})\nonumber\\
&& \dots (1+f_{N-1,N})
\biggr]d^{3}r_{1}d^{3}r_{2}\dots d^{3}r_{N}.\label{q2}
\end{eqnarray}
For point masses, the Hamiltonian and, hence, the partition function diverge  at $r_{ij}=0$. 
This divergence has been removed by taking the extended nature of particles into account, via 
introducing a softening parameter which takes care of the finite size of each galaxy. Thus,  
by 
introducing the softening parameter, the Newtonian (interaction) potential energy between particles is given  by
\begin{equation}
\Phi(r_{ij})=-\frac{Gm^2}{(r_{ij}^2+\epsilon^2)^{1/2}},
\end{equation}
where $\epsilon$ represents the softening parameter.

In the cosmological constant $\Lambda$       model, where the galaxies interact 
via Newtonian potential, the   potential energy  with the softening parameter, 
is given by \citet{pe}
\begin{equation}
\Phi(r_{ij})=-\frac{Gm^2}{(r_{ij}^2+\epsilon^2)^{1/2}}-\frac{\Lambda r_{ij}^2}{6}.
\end{equation}
Here, we do not introduce the softening parameter $\epsilon$ in the second term of right hand side as corresponding Hamiltonian does not diverges.

Now, the expression of two particle function  (\ref{fun}) for above potential energy
reads,
\begin{equation}
f_{ij}=exp\biggl[\frac{Gm^{2}}{(r_{ij}^{2}+\epsilon^{2})^{1/2}T}+\frac{\Lambda r_{ij}^2}{6T}\biggr]-1.
\end{equation}
 Assuming moderately dilute systems, the two particle function $f_{ij}$ upon  expansion  gives
\begin{eqnarray}
f_{ij}=\biggl[\frac{Gm^{2}}{(r_{ij}^{2}+\epsilon^{2})^{1/2}T}+\frac{\Lambda r_{ij}^2}{6T}\biggr].
\end{eqnarray}
Evaluating the configuration integrals over a spherical volume of radius $R_{1}$, which utilize equations (\ref{q1}) and (\ref{q2}), gives 
\begin{equation}
Q_{1}(T,V)=V.
\end{equation}
Now, configuration integral $Q_{2}(T,V)$ can be written as
\begin{equation}
Q_{2}(T,V)=4\pi V \int_{0}^{R_{1}}\left[1+\frac{Gm^2}{T(r^{2}+\epsilon^{2})^{1/2}}+\frac{\Lambda r^2}{6T}\right]r^2dr.
\end{equation}
Upon performing integration,  this yields
\begin{eqnarray}
Q_{2}(T,V)&=&V^2\left[1+ \frac{\Lambda R^2_1}{10T} +3 \frac{Gm^2}{R_1T}\left(\frac{1}{2}\sqrt{1+\frac{\epsilon^2}{R_1^2}}\right.\right.\nonumber\\
&+&\left.\left. \frac{1}{2}\frac{\epsilon^2}{R_1^2} \log \frac{\epsilon/R_1}{\left[ 1+\sqrt{1+\frac{\epsilon^2}{R_1^2}}\right]}\right) \right].
\end{eqnarray}
With the help of following definitions:  
\begin{eqnarray}
&&\alpha_1\left(\frac{\epsilon}{R_1}\right) = \sqrt{1+\frac{\epsilon^2}{R_1^2}} +\frac{\epsilon^2}{R_1^2} \log \frac{{\epsilon/R_1}}{\left[ 1+\sqrt{1+\frac{\epsilon^2}{R_1^2}}\right]},\nonumber\\
&& 
  \alpha_2=\frac{\Lambda R_1^3}{15 Gm^2}.
\end{eqnarray}
The above expression $Q_{2}(T,V)$ can further be written in compact form as,
\begin{equation}
Q_{2}(T,V)=V^2\biggl(1+ \frac{3}{2}\left(\alpha_1+ {\alpha_2} \right)\left(\frac{ Gm^{2}}{ R_{1}T}\right)^3\biggr).\label{a}
\end{equation}
Here we have utilized the scale transformations  $\rho\rightarrow \lambda^{-3}\rho, T\rightarrow\lambda^{-1}T$ and $R_1\rightarrow \lambda R_1$ also,
to transform $\frac{Gm^2}{R_1T}\rightarrow (\frac{Gm^2}{R_1T})^3$.
Since $R_{1}\sim \rho^{-1/3}\sim (\bar N/V)^{-1/3}$, so we can write 
\begin{equation}
  \frac{3}{2}\left(\frac{ Gm^{2}}{ R_{1}T}\right)^3=\frac{3}{2}\left(\frac{ Gm^{2}}{  T}\right)^3\rho := x.\label{sc}
\end{equation}
Thus, we can write equation (\ref{a}) as:
\begin{equation}
Q_{2}(T,V)=V^2\big(1+\alpha x\big),
\end{equation}
where 
\begin{equation}
\alpha\left(\frac{\epsilon}{R_1}\right) =\alpha_1\left(\frac{\epsilon}{R_1}\right)+ {\alpha_2}.
\end{equation}
For the point masses  (i.e., $\epsilon =0$), $\alpha$ reduces to
\begin{equation}
\alpha\left( \epsilon =0\right) =  1 +{\alpha_2}.
\end{equation}
Following similar procedure, we obtain configurational integral for higher orders as:
\begin{equation}
Q_{3}(T,V)=V^3\big(1+\alpha x\big)^{2},
\end{equation}
and
\begin{equation}
Q_{4}(T,V)=V^4\big(1+\alpha x\big)^{3}.
\end{equation}
Thus, for most general case,  we have
\begin{equation}
Q_{N}(T,V)=V^N\big(1+\alpha x\big)^{N-1}.\label{qn}
\end{equation}
Hence, the gravitational partition function is obtained explicitly by substituting the value
 of $Q_{N}(T,V)$ given in (\ref{qn}) to (\ref{zn}) as,
\begin{equation}
Z_N(T,V)=\frac{1}{N!}\left(\frac{2\pi mT}{\lambda^2}\right)^{3N/2}V^{N} (1+\alpha x)^{N-1}.
\end{equation}
Here, the  effect of cosmological constant, embedded in $\alpha$, can be seen in the expression of the gravitational partition function.
\section{Equations of State}
We compute below the various thermodynamic quantities relevant for strongly interacting system of galaxies interacting through a   
Newtonian potential describing galaxies  interacting with each other .  
It is well-known that the thermodynamic quantities can  be easily calculated from the gravitational partition function.
For example, Helmholtz free energy, defined, generally, by
$F=-T\ln Z_{N}(T,V)$ is calculated as
\begin{equation}
F=-T\ln\biggl(\frac{1}{N!}\left(\frac{2\pi mT}{\lambda^2}\right)^{3N/2}V^N\big(1+\alpha x\big)^{N-1}\biggr). \label{f}
\end{equation}
Further simplification leads to
\begin{eqnarray}
F&=&NT\ln\left(\frac{N}{V}T^{-3/2}\right)-NT -(N-1)T\ln\big(1+\alpha x\big)
\nonumber\\
& -&\frac{3}{2}NT\ln\left(\frac{2\pi m}{\lambda^2}\right). 
\end{eqnarray}
Now, it is easy to compute entropy $S$ for a given Helmholtz free energy with formula,
$S= -\biggl(\frac{\partial F}{\partial T}\biggr)_{N,V}$. 
Here entropy reads
\begin{eqnarray}
S&=&N\ln\left(\frac{V}{N}T^{3/2}\right)+(N-1)\ln\big(1+\alpha x\big)\nonumber\\
&-&3N\frac{\alpha x}{1+\alpha x}+\frac{5}{2}N+\frac{3}{2}N\ln\left(\frac{2\pi m}{\lambda^2}\right).\label{s}
\end{eqnarray}
 For large $N$ such that $N-1\approx N$, 
this  can further be simplified as
\begin{equation}
\frac{S}{N}=\ln\left(\frac{V}{N}T^{3/2}\right)-\ln\big(1-\frac{\alpha x}{1+\alpha x}\big)-3\frac{\alpha x}{1+\alpha x}+\frac{S_{0}}{N},
\end{equation}
where  definition $S_{0}=\frac{5}{2}N+\frac{3}{2}N\ln\left(\frac{2\pi m}{\lambda^2}\right)$ is
 utilized. 
 Comparing this expression  to its  standard form \citet{ahm02}, the  clustering parameter   
   of galaxies  in the expanding universe,  $\mathcal{B}$, is derived as   by 
\begin{eqnarray}
\mathcal{B}=\frac{\alpha x}{1+\alpha x}.
\label{b}
\end{eqnarray}
Evaluating the clustering parameter is worth  because it plays a crucial role in finding   various thermodynamic quantities. 
For the point masses,   $\epsilon =0$,  the  clustering parameter  in modified potential becomes
\begin{eqnarray}
\mathcal{B}(\epsilon =0):=\mathcal{B}_0  =\frac{(1+\alpha_2) x}{1+x(1+\alpha_2)}.
\end{eqnarray}
Thus, due to cosmological constant modified potential, the  clustering parameter for the point masses get following correction:
 \begin{eqnarray}
\mathcal{B}_0  =  b \left( \frac{1 +\alpha_2 }{ 1 +b\alpha_2  } \right),
\end{eqnarray}
where   $b= {x}/\left[{1+ x}\right]$ \citet{ahm02} refers the original clustering parameter of  Newtonian potential for point masses.

Employing expression (\ref{f}) and (\ref{s}),  the internal energy, defined as $U =  F+TS$, for a system of galaxies is calculated by
\begin{equation}
U=\frac{3}{2}NT\big(1-2\mathcal{B}\big).
\end{equation}
It is evident from the above expression that the internal energy depends on the 
cosmological constant embedded in clustering parameter $\mathcal{B}$.

The   pressure    and chemical potential, utilizing the standard notations and definitions
$P= -\left(\frac{\partial F}{\partial V}\right)_{N,T}$ and  $\mu = \biggl(\frac{\partial F}{\partial N}\biggr)_{V,T}$, respectively, are calculated by 
 \begin{eqnarray}
  P&=&\frac{NT}{V}\big(1-\mathcal{B}\big),\label{p}\\
  {\mu}&=&{T}\ln\left(\frac{N}{V} T^{-3/2}\right)+{T}\ln\big(1-\mathcal{B}\big)\nonumber\\
  &-&\frac{3}{2}{T}\ln\left(\frac{2\pi m}{\lambda^2}\right)-\mathcal{B}{T}.
 \end{eqnarray}
 Here, we observe that the  pressure    and chemical potential also get correction
 due to the cosmological constant as clustering parameter depends on cosmological constant  explicitly. 
\section{General Distribution Function}

The definition of grand canonical partition function is,
\begin{eqnarray}
Z_{G}(T,V,z)=\sum_{N=0}^{\infty}z^NZ_{N}(V,T), \label{g}
\end{eqnarray}
where $z$ is the activity. 
The  grand partition function for our  gravitationally interacting system of galaxies is 
calculated by
\begin{equation}
\ln Z_{G}=\frac{PV}{T}=  N(1-\mathcal{B}),\label{z}
\end{equation}
where the expression (\ref{p}) is
 utilized.

Now, the probability of finding $N$ particles in volume $V$ can be 
 estimated by relation
\begin{eqnarray}
F(N)=\frac{\sum_{i}e^{\frac{N\mu}{T}}e^{\frac{-U_i}{T}}}{Z_{G}(T,V,z)}=\frac{e^{\frac{N\mu}{T}}Z_{N}(V,T)}{Z_{G}(T,V,z)}.
\end{eqnarray}
Here, with the help of (\ref{z}), the distribution function for a system of point masses is computed, precisely, by
\begin{eqnarray}
&&F(N,\epsilon =0)=\frac{\bar{N}^{N}}{N!}\biggl(1+\frac{N}{\bar N}\frac{\mathcal{B}_0}{(1-\mathcal{B}_0)}\biggr)^{N-1}
\nonumber\\
&&\times \biggl(1+\frac{\mathcal{B}_0}{(1-\mathcal{B}_0)}\biggr)^{-N}e^{[-N\mathcal{B}_0 -\bar N(1-\mathcal{B}_0)]}.
\end{eqnarray}
In the similar fashion, the distribution function for non-point mass
particles is derived as
\begin{eqnarray}
&&F(N,\epsilon)=\frac{\bar{N}^{N}}{N!}\biggl(1+\frac{N}{\bar N}\frac{\mathcal{B}}{(1-\mathcal{B})}\biggr)^{N-1}\nonumber\\
&&\times\biggl(1+\frac{\mathcal{B}}{(1-\mathcal{B})}\biggr)^{-N}e^{[-N\mathcal{B} -\bar N(1-\mathcal{B})]}.
\end{eqnarray}
Remarkably, the structure of resulting   distribution function coincides exactly with the derived earlier \citet{ahm02}. The only difference here we 
get, the cosmological constant corrected clustering parameter $\mathcal{B}$ in place of
usual clustering parameter $b$.

\section{Discussion and Conclusion}
 \begin{figure}[htb]
 \begin{center}
 \begin{tabular}{cc}
\rotatebox{270}{\resizebox{60mm}{!}{ \includegraphics[width=150pt]{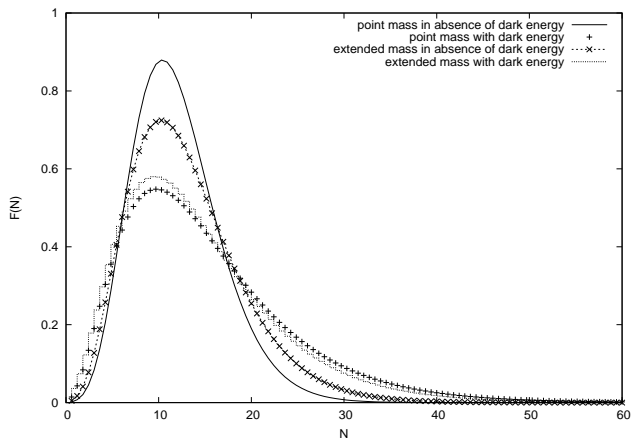}}}
\end{tabular}
 \end{center}
 \caption{Comparative study of distribution function $F(N)$ with 
 and  without dark energy corrections for $b=0.3$, $\bar{N}=10$. Here for  point mass
   $\alpha_1 =1$ with dark energy  (i.e. $\alpha_2 =1$) and without dark energy  (i.e. $\alpha_2 =0$) and  for extended mass $\alpha_1 =0.7$ with dark energy  
   (i.e. $\alpha_2 =1$) and without dark energy  (i.e. $\alpha_2 =0$).}
 \end{figure}

In this section, we draw a comparative analysis to emphasize the effect of 
dark energy corrections on the distribution function. Here, in figure 1, we consider both the cases 
of  point mass in absence and presence of dark energy and the
extended mass in absence and  presence of dark energy. 
The point mass in absence of dark energy has the highest value peak  $F(N)$, 
and the point mass with dark energy has the lowest value peak value of $F(N)$. 
The extended mass has a softening parameter, and so the peak values of extended mass
for both these cases lies in between these two values. The peak value of 
the extended mass with dark energy is less than the peak value of extended mass 
in absence of dark energy. It is also observed that $F(N)$ for
the point mass in absence 
of dark energy reduces faster than all the other cases. Thus, its value 
becomes lower than all the other cases, at  
later stages. This distribution can be used to compare the predicted values 
of $F(N)$, with the observed values and thus used to check the validity of this 
analysis. 

In this paper, we have considered a  strongly interacting system of galaxies in the
expanding universe and derived,  utilizing the configuration integrals over a spherical volume,   the gravitational partition function for galaxies
interacting with each other. 
The cosmological constant  correction is evident in the expression obtained for gravitational partition function. We have computed various 
thermodynamical quantities, for example, 
 Helmholtz free energy, entropy, internal energy, pressure and chemical potential, which depend on cosmological constant
 modified clustering parameter explicitly, to study the exact equations of state for galaxies. The  cosmological constant modified
 clustering parameter is compared with the original clustering parameter obtained for Newtonian potential. 
 Further, we have studied  the general distribution function for such system.
 We would like to point out that unlike the Jacobson formalism \citet{jac}, in this paper,  the fundamental laws of gravity are not 
obtained using thermodynamics, but only the thermodynamical consequences of a system of galaxies are analysed. 

The modification in the gravity action can affect the gravitational potential in the low 
energy limit and the modified potential reduces 
to the Newtonian one on the solar system scale as well. It has been seen in 
\citet{[32],[33],[330],[34]} that the
modified gravitational potential could fit galaxy rotation curves without considering dark matter or dark energy. In fact, this provides an opportunity to draw a formal
analogy between the corrections due to the modified Newtonian potential and the dark energy models. In general, a relativistic gravity theory leads to a
change to the Newton potential  \citet{[36]}. In
the post-Newtonian  formalism, this could provide tests for the  theory 
\citet{[24],[37],[38],[39]}.
Verlinde's derivation of laws of gravitation provides a new direction to understand gravity
from the first principles. 
The  entropic
force interpretation gets relevance  in various contexts \citet{45,46,47,48,49,50,51,52,53,54}. 
The Friedmann
equations governing the dynamical evolution of the FRW
universe from the viewpoint of entropic force together with
the equipartition law of energy and the Unruh temperature is advocated \citet{34,
35}.     Corrections to Newton's law of gravitation as well as modified Friedmann equations from the 
entropic force point of view are discussed in \citet{36, 37}. Further   entropic interpretation of gravity 
has also been used to study the modified Newton's law \citet{38}, the Newtonian
gravity in loop quantum gravity \citet{39}, the holographic dark energy \citet{40,41,42} thermodynamics
of black holes \citet{43} and the extension to Coulomb force \citet{44}. It may be noted that it is possible to study 
the clustering of galaxies using this modified Newton's law. In fact, Newton's law also gets modified 
due to brane world effects, and the 
 clustering parameter of galaxies has been studying using such a modified  Newton's law \citet{mir}.
 Thus, it would be interesting to analyse the effect of modified Newton's law obtained from entropic force on the 
 clustering of galaxies.

It may be noted that the clustering of galaxies has also been studied using the modified Newtonian potential produced by $f(R)$ gravity \citet{1aa,2aa}. 
In fact these models can explain the dynamics of spiral and elliptical galaxies, even in absence of dark matter. 
This analysis was performed through the pressure profile of galaxy clusters. It was assumed that 
this   is in hydrodynamic equilibrium within the potential well of the modified gravitational potential.
Furthermore, it was demonstrated that this model is consistent with Planck data. We expect a similar effect to occur, 
if we perform this analysis by modifying the gravitational field by the cosmological constant. This is because the cosmological constant can be obtained from a suitable $f(R)$ gravity model \citet{4aa}. Now as it has been demonstrated that the $f(R)$ gravity is consistent with Planck data \citet{1aa,2aa}, we expect the model studied in this paper to also be consistent with the Planck data. This can be observed from the fact that both the modified Newtonian constant and the cosmological constant term will modify the gravitational potential energy term in the partition function, and hence, we expect on physical grounds that they will produce similar results. We would also like to comment that unlike the   approach where the distribution of galaxies was studied using the  standard Boltzmann-Vlasov equation, in our approach we analyses the thermodynamical properties of galaxies. Just like the other approaches, here, the galaxies are approximated as point particles, but calculating thermodynamic quantities for such systems makes it easier to make physical predication using this approach. We would also like to comment that as $f(R)$  has been used for analyzing galaxies clustering by analyzing the  effect of such a modification of Newtonian gravitational potential on the distribution function of galaxies, it would be interesting to use this modified Newtonian gravitational potential \citet{1aa,2aa} in the thermodynamic approach. The advantage of using this  approach is that it can be used for analyzing the gravitational phase transition \citet{5aa}, and such a phase transition can also be studied for the model proposed in this paper, and the modified Newtonian potential by $f(R)$ gravity.

\end{document}